\newcommand{\be}{\begin{equation}}
\newcommand{\ee}{\end{equation}}
\newcommand{\bea}{\begin{eqnarray}}
\newcommand{\eea}{\end{eqnarray}}
\newcommand{\del}{\nabla}
\newcommand{\delc}{\overline{\nabla}}
\documentstyle[preprint,revtex,eqsecnum]{aps}
\begin{document}
\draft
\preprint{Alberta-Thy-24-92}
\preprint{gr-qc/9211016}
\begin{title}
VACUUM POLARISATION AND THE BLACK HOLE \\SINGULARITY
\end{title}
\author{Warren G. Anderson, Patrick R. Brady, \\and Roberto Camporesi}%
\begin{instit}
Theoretical Physics Institute, University of  Alberta,\\
Edmonton, Alberta, Canada T6G 2J1
\end{instit}
%
%

\begin{abstract}
In order to investigate the effects of vacuum polarisation on mass inflation
singularities, we study a simple toy model of a charged black hole with cross
flowing radial null dust which is homogeneous in the black hole interior. In
the region $r^2 \ll e^2$ we find an approximate analytic solution to the
classical field equations. The renormalized stress-energy tensor is evaluated
on this background and we  find the vacuum polarisation backreaction
corrections to the mass function $m(r)$. Asymptotic analysis of the
semiclassical mass function shows that the mass inflation singularity is much
stronger in the presence of vacuum polarisation than in the classical case.
\end{abstract}
\pacs{PACS numbers: 97.60Lf, 04.60+n\\
{To appear: Class. Quantum Grav.}}
\clearpage

\narrowtext
%
%

\section{Introduction}  \label{sec:intro}
The causal structure of charged or rotating black holes in General
Relativity~(GR) suggests the possibility of travelling through such a hole to
another universe.
However a closer look soon reveals a difficulty which was first pointed out by
Penrose~\cite{Pen:68}; the inner (Cauchy) horizon is unstable due to the
infinite blueshift of infalling radiation. In  a generic collapse such
radiation will always be present, due to the tail of backscattered radiation
{}from the body which produced the black hole.  The divergence of linear
perturbations on this horizon has been verified by several
authors~\cite{Pen:73}, and it was recently  shown that this blueshifted
radiation combined with outflow {}from the star  produces a scalar curvature
singularity. Since quantitative results exist only in the spherical case (i.e.
for charged black holes), attention is restricted to these spacetimes in what
follows.

Hiscock~\cite{His:81} has modelled the backscattered radiation as a null dust
and shown that  this  results in an observer dependent singularity at the inner
horizon. This is a ``whimper'' singularity, in the classification scheme of
Ellis and King~\cite{EllisK}, and none of the curvature invariants diverge.
However, whimper singularities are generally believed to be unstable to
perturbations~\cite{Sik:79}. One such perturbation to Hiscock's model, which
preserves the spherical symmetry, is the addition of a radial outflowing stream
of null dust (which provides a crude way to model radiation {}from the
collapsing
star). Poisson and Israel~\cite{P&I:89} have shown that for such a crossflow, a
phenomenon which has come to be known as ``mass inflation'' occurs. The mass
function of the spherical black hole becomes infinite on the inner horizon
causing a scalar curvature singularity to occur.

More recently, Ori~\cite{Ori:91} has introduced a model in which the outflow
is compressed into a thin null shell, for which he  finds an exact mass
inflation solution to the Einstein-Maxwell equations. His analysis of the
resulting mass inflation singularity leads him to conclude that, although it is
stronger than the ``whimper'' singularity of the purely infalling null
radiation, it is still rather weak in the sense that a physical body
approaching it will experience only a finite deformation due to gravitational
tidal forces. This has lead to intense debate about the possible continuation
of spacetime beyond this null singularity~\cite{H&H:92}.

It is now generally believed, however, that the appearance of  singularities
in classical GR signal regimes in which a quantum theory of gravity is needed
for a correct physical description. Although there exists no satisfactory
quantum theory of gravity at present, a semiclassical approximation may be
useful for investigating the behavior of spacetimes in regions of high
curvature. Our purpose in this paper is to investigate the mass inflation
singularity using the machinery of semiclassical gravity.

For simplicity, we will use as our classical mass inflation background a toy
model developed by Page  and independently by Ori~\cite{Pag:92}. This model
(henceforth known as the homogeneous mass inflation or HMI model) consists of a
spherically symmetric charged black hole interior with crossflowing null
radiation which is homogeneous in the region between the inner and outer
horizons. The HMI model has the advantage that the unknown metric functions
depend only on the radial  coordinate $r$. Furthermore, it is important to note
that Page has shown that in this model the mass inflation singularity is not
null but rather spacelike, and it is tempting to speculate that this will also
be true of  part of the singularity in more realistic models.

In  section~\ref{sec:HMIclassical} we will present a classical solution for
the HMI model in the region $r^2 \ll e^2$ where $e$ is the charge of the black
hole. This particular solution did not explicitly appear in the work of Page
and Ori~\cite{Pag:92}, although it is implicit therein.  In
Section~\ref{sec:VP}, we will find the renormalised stress-energy tensor for a
conformally coupled massless scalar field on the classical HMI background, and
we argue that the dominant terms should be insensitive to the quantum state we
use.  The backreaction equations are then solved approximately, to find vacuum
polarisation corrections to the classical background. In particular it is shown
that the correction to the mass function diverges more strongly than in the
classical case.

%
%

\section{The HMI Background}      \label{sec:HMIclassical}

The spherically symmetric line element may always be written in the form
\be
ds^2 = (\frac{r^2}{e^2}) [ d\sigma^2 + e^2d \Omega^2],\qquad
d\sigma^2 = \gamma_{ab}dx^a dx^b,       \label{eq:metric}
\ee
where $d\Omega^2$ is the line element on a unit sphere, and $e$ is a constant
with the dimensions of length. Latin indices $a,b,\ldots$ range over (0,1).
Thus the problem reduces to finding the two dimensional metric $\gamma_{ab}$,
and the scalar function $r(x^a)$ which is defined so that the area of a two
sphere is $4\pi r^2$.  We introduce the scalar invariant
\be
\gamma^{ab} r_{,a} r_{,b} = -e^{-2} B(r)
\equiv - e^{-2} ( 2 m(x^a) r - r^2 -	  e^2 ),
\label{eq:Bdefn}
\ee
where $m(x^a)$ is the mass of the black hole interior. It is convenient to use
$r(x^a)$ as a coordinate so that $d\sigma^2 = -e^2dr^2/B(r) + g(r)dt^2$.

The Einstein field equations  can now be written in the covariant form
\bea
2  e^2 r \gamma^{ab} r_{;ab} &=&  4 m(r) r - 4 e^2 + 8 \pi r^2 e^2
\gamma^{ab} T_{ab},  \label{eq:EFE1}   \\
m_{,a} &=& 4 \pi  e^2[  \gamma^{cb}
T_{ac}- \delta^b{_a} \gamma^{cd}T_{cd} ]r_{,b},     \label{eq:EFE2}
\eea
 where the semicolon (;) denotes the covariant derivative associated with
$d\sigma^2$, and  $T_{ab}$ is the $2\times 2$  submatrix of the stress-energy
tensor.

The stress-energy tensor for a homogeneous crossflow of null radiation takes
the form
\be
-T{^t}_t = T{^r}_r = \rho (r)\: ,\ \ \  T_{tr} = X(r).
\label{eq:Tabdefn}
\ee
Covariant conservation of the stress tensor requires that  $\rho(r) =
-C^2/(g(r) r^4)$ and $X(r) = D(r^4B(r)g(r))^{-1/2}$.
Without loss of generality, we will take the constants of integration to be
$C^2= -|e|D/2 = {e^2}/{4 \pi}$, since any arbitrariness can in fact be absorbed
into the coordinate $t$.

We are now in a position to solve the classical equations for which  $T{^a}_a =
0$ and $T^t{_t} = {e^2}/({4 \pi g(r) r^4})$. We can reduce the system
(\ref{eq:EFE1}) and (\ref{eq:EFE2}) to a single integral equation for the mass
function
\be
\frac{d m(r)}{dr} = \frac{r^2}{B(r)} \int \frac{4}{r} \left( \frac{d\;
m(r)}{dr}
\right)^2 dr\: .
\ee
Since we are interested in the behavior near the mass inflation singularity,
which Page \cite{Pag:92}  has shown to be at $r = 0$, we will consider only the
region $r^2 \ll e^2$, so that $B(r) \approx 2 m(r)\; r - e^2$.  With this
simplifying assumption it is easily verified that
\be
m(r) = \frac{e^2}{r}, \qquad g(r) =  {1},     \label{eq:gdefn}
\ee
is a solution to the field equations. Thus we have the HMI background metric
\be
ds^2 = \frac{r^2}{e^2} [ -dr^2 + dt^2 + e^2 d \Omega^2]\: ,
\label{eq:HMImet}
\ee
{}from which it is clear that the singularity is spacelike.
This simple form will allow us to analyse the vacuum polarisation in some
detail.  Note that this is not the general solution to the equations, about
which Page and Ori~\cite{Pag:92} were able to obtain qualitative information.
However the main feature, that the mass function inflates to infinity like
$1/r$, is captured in our solution.

%
%

\section{Vacuum Polarisation}      \label{sec:VP}
Next, we wish to consider semiclassical corrections to the metric functions,
due to the presence of a conformally coupled massless scalar field $\phi$
obeying
\be
{g}^{\alpha \beta} \del_{\alpha} \del_{\beta} \phi - \frac{1}{6}
{R} \phi = 0\: .
\label{eq:confscalwav}
\ee
Later we will argue that our conclusions should be insensitive to the quantum
state in which we evaluate the stress energy tensor. For now we ignore this
problem and present a derivation for states possessing the natural symmetries
of the manifold; $R^2 \times S^2$.  Our method is to calculate the
$\zeta$-function and effective potential for $d\overline{s}^2 = -dr^2 + dt^2 +
e^2d\Omega^2$, and then to use Page's generalisation of a result due to  Brown
and Cassidy to obtain the stress energy tensor in the physical spacetime.

Some detail is included below for completeness. The $\zeta$-function is
obtained {}from the heat kernel for this space (which is homogeneous) and is
defined via the Mellin transform
\be
\zeta(s) = \frac{1}{ \Gamma(s)}\int_0^{\infty}
t^{s-1}  K(0,0;t) dt.
\ee
Here
\bea
K(0,0;t)&=& ((4\pi\;e)^2\;t)^{-1}\sum^{\infty}_{l=0}(2l+1)\;
\exp[-t(12(l+1/2)^2 + 1)/12e^2]
\eea
is the heat kernel for $d\overline{s}^2$ at $x=x'=0$.  After analytically
continuing this $\zeta$-function to a meromorphic function in the complex
$s$-plane we find
\be
\zeta(s) = \frac{2 e^{2s-4}}{(4\pi)^2}
\sum^{\infty}_{k=0} \frac{\Gamma(s+k-1) (-1)^k}{\Gamma(s) k! 12^k}
\zeta_{R}(2s+2k-3,\; 1/2)  \label{eq:zetas}
\ee
where $\zeta_{R}(z,q)$ is the Riemann-Hurwitz zeta  function~\cite{Grad&R}. The
effective potential is formally given by $V = (-1/2)[{\zeta '(0) + \log(\mu^2)
\zeta(0)}]$~\cite{B&D:82} and is now easily evaluated using (\ref{eq:zetas}).
It is
\bea
V &=& - \left(\frac{A}{16\pi e^4} + \frac{T}{2}\;\ln(e^2 \mu^2)\right),
\label{eq:effpotconf} \\
A &=&  \sum_{k=3}^{\infty}
\frac{(-1)^k}{k(k-1)\,12^k } \zeta_R(2k-3,1/2)
-2\zeta '_R(-3,1/2)\nonumber \\&&-\frac{1}{6}\zeta
'_R(-1,1/2)+\frac{7}{960}-\frac{1}{288}\psi(1/2),
\eea
where $\mu$ is an arbitrary mass scale,  $	T = (1440 \pi^2 e^4)^{-1}$ is the
trace anomaly, $\zeta '_R(z,q)\equiv \partial \zeta_R(z,q)/\partial z$, and
$\psi(x)=\Gamma '(x)/\Gamma(x)$. By performing the usual variation of the
effective action we obtain the stress tensor in this static space
\bea
\bigl<\overline{T}^\alpha{_\beta}\bigr>_{ren} &=&
\frac{2\overline{g}^{\alpha\mu}}{\sqrt{\overline{g}}}
\frac{\delta\left( \sqrt{\overline{g}} \: V\right)}{\delta
\overline{g}^{\beta\mu}} =  \mbox{diag}[-V,-V,V+\frac{T}{2},V+\frac{T}{2}].
\label{eq:TQconf}
\eea
 In fact, one may evaluate the above stress-energy tensor for massless
conformally coupled fields of other spins.  This only changes the numerical
values of the $T$ and $A$ above. For spin-$1/2$ (two-component) theory  the
trace anomaly is $T=(480\pi^2 e^4)^{-1}$ and
\be
A = 4\zeta '_R(-3)+\frac{1}{60},
\ee
where $\zeta_R(z)$ is the Riemann zeta function~\cite{Grad&R}. For the Maxwell
field we find $T=(120\pi^2e^4)^{-1}$ and
\bea
A &=&  \sum_{k=3}^{\infty}\frac{1}{k(k-1)\,2^{2k-1} }
\zeta_R(2k-3,3/2)
-4\zeta '_R(-3,3/2)\nonumber \\ &&+ \zeta
'_R(-1,3/2)+\frac{127}{480}-\frac{1}{16}\psi(3/2).
\eea

We now invoke a result due to Page~\cite{Pag:82}. He finds that  the
renormalised stress-energy tensor for a conformally coupled massless scalar
field transforms conformally as
\bea
T^\mu{_ \nu}  &=& \Omega^{-4} \overline{T}^\mu{_\nu}  - 8 \alpha
\Omega^{-4}[ \delc_\alpha
\delc^\beta (\overline{C}^{\alpha \mu}{_{\beta \nu}} \ln \Omega) +
\frac{1}{2}
\overline{R}_\alpha{^\beta} \overline{C}^{\alpha \mu}{_{\beta \nu}}
\ln \Omega]  \nonumber\\
&&	+ \beta[(R_\alpha{^\beta} C^{\alpha \mu}{_{\beta \nu}} - 2
H^\mu{_\nu})-
\Omega^{-4}(\overline{R}_\alpha{^\beta} \overline{C}^{\alpha
\mu}{_{\beta \nu}} - 2
\overline{H}^\mu{_\nu})]
\nonumber\\
&&- \frac{1}{6} \gamma [I^\mu{_\nu} - \Omega^{-4}
\overline{I}^\mu{_\nu}], \label{eq:Tconftran}
\eea
where
\bea
H_{\mu \nu} &\equiv& - R^\alpha{_\mu} R_{\alpha \nu} + \frac{2}{3} R
R_{\mu \nu}
+ (\frac{1}{2}
R^{\alpha}{_\beta} R^\beta{_\alpha} - \frac{1}{4} R^2) g_{\mu
\nu},    \label{eq:Hdefn} \\
I_{\mu \nu} &\equiv& 2 R_{| \mu \nu} - 2 R R_{\mu \nu} + (\frac{1}{2}
R^2 - 2 R_{|
\alpha}{^\alpha}) g_{\mu \nu},     \label{eq:Idefn}
\eea
and, in the case of a scalar field, $\alpha = {12}/{(2^9 45 \pi^2)}$,	$\beta =
{-4}/{(2^9  45 \pi^2)}$,	$\gamma = {8}/{(2^9 45 \pi^2)}$. The barred tensors
are evaluated in $d\overline{s}^2= \Omega^{-2} ds^2$, and in our case $\Omega^2
= r^2/e^2$.  It is now a straightforward matter to obtain
$\bigl<T^\alpha{_\beta}\bigr>_{ren}$ for the metric (\ref{eq:HMImet}). We are
interested, however, only in the effect of $\bigl<T^\alpha{_\beta}\bigr>_{ren}$
near the mass inflation singularity. Thus, we present here only the dominant
contributions in the asymptotic limit $r \rightarrow 0$. The relevant
contributions are
\bea
\bigl<T^t{_t}\bigr>_{ren}  &\sim& - 10 \beta e^4 r^{-8},\label{eq:TQtt}
\\
\bigl<T^r{_r}\bigr>_{ren}  &\sim& 6 \beta e^4 r^{-8}.
\label{eq:TQrr}
\eea
It is worth noting that for the spin 1/2 and spin 1 fields the dominant
contributions are identical apart {}from a positive multiplicative constant.

Before examining the backreaction, some comments are in order about the
quantum state in which this stress energy tensor is evaluated.  The boundary
conditions defining the state in which our system was prepared should appear in
the derivation of $\bigl< \overline{T}^\mu{_\nu}\bigr>$.  However, it is not an
easy task to impose these in this case, since we only have an asymptotic
classical solution for $r^2 \ll e^2$.  Instead we suggest that the asymptotic
behavior will be insensitive to the initial conditions.  In particular, since
the singularity is spacelike, one may ignore the quantum influx due to Hawking
radiation as it will most likely be swamped by the classical radiation.  Other
non-local effects are expected to contribute to the full stress-energy tensor
as $\int \sqrt{-g} (curvature)^2 dr$ \cite{H&W:80}; thus these effects will
behave at most like $r^{-3}$.  For these reasons we feel that  the dominant
behavior is captured with our present model.

%
%

\section{Backreaction}     \label{sec:BR}
To solve for the backreaction of the vacuum polarisation due to (\ref{eq:TQtt})
and (\ref{eq:TQrr}) on the HMI background, we consider the following
expansions:
\bea
m(r) &=& m^{(0)} (r) + \hbar m^{(1)}(r) + O(\hbar^2),
\label{eq:mexp} \\
g(r) &=& g^{(0)}(r) + \hbar g^{(1)}(r) + O(\hbar^2),
\label{eq:gexp} \\
T_{ab} &=& T^{(0)}_{ab} + \hbar \bigl<T_{ab}\bigr>_{ren} + O(\hbar^2)
\label{eq:Texp}
\eea
where $m^{(0)} (r)$, $g^{(0)}(r)$, and $T^{(0)}_{ab} $ are the classical metric
functions and stress-energy tensor given by (\ref{eq:gdefn}) and
(\ref{eq:Tabdefn}) respectively.

Substituting (\ref{eq:mexp}) - (\ref{eq:Texp}) into the field equations
(\ref{eq:EFE1}) and (\ref{eq:EFE2}) and keeping only terms of less than
$O(\hbar^2)$ we obtain the backreaction equations for the terms of  $O(\hbar)$
\bea
e^2 \frac{d g^{(1)}}{dr} + 2 r \frac{d m^{(1)}}{dr} + 6 m^{(1)}
&=& -8\pi r^3 \bigl<T^a{_a}\bigr>_{ren} \: ,
\label{eq:QEFE1} \\
-e^2 g^{(1)} + r^2 \frac{d m^{(1)}}{dr}
&=& - 4 \pi r^4 \bigl<T^t{_t}\bigr>_{ren}\: .   \label{eq:QEFE2}
\eea
Eliminating $g^{(1)}$ {}from these equations and inserting only the dominant
terms (\ref{eq:TQtt}) and (\ref{eq:TQrr}) for the $\bigl<T^a{_b}\bigr>_{ren}$
contributions we obtain
\be
r^2 \frac{d^2 m^{(1)}}{dr^2} + 4 r \frac{d m^{(1)}}{dr} + 6 m^{(1)}
\simeq - \frac{128 \pi \beta e^4}{r^5},
\label{eq:backreaceq}
\ee
which is valid as $r\rightarrow 0$.  A particular solution to this equation is
\be
m^{(1)} \simeq -\frac{8 \pi \beta e^4}{r^5}.
\ee
Recalling  that $\beta$ is negative, we see that $m^{(1)}$ diverges even more
strongly than the classical mass function ($m^{(0)} \sim r^{-1}$).

%
%

\section{Conclusion}
It is widely believed that singularities in GR indicate regimes where a
quantum treatment is necessary for a correct physical description. We have
attempted a semiclassical analysis of the region near the mass inflation
singularity for the  HMI model. This is only a toy model which is useful as a
tool to begin our investigations.  The main feature is the spacelike
singularity which is present, in contrast to the other mass inflation models
which have a null singularity.  In our opinion, it is entirely possible that
the singularity inside a generic black hole is  spacelike, however, this
present model is too simple to be considered as realistic enough to describe
such a situation.   Our findings indicate that in the region of the mass
inflation singularity, vacuum polarisation effects  lead to a worsening of the
singularity, rather  than having a regulating effect as might be hoped. If this
remains true in more realistic analyses, we will have to wait for a theory of
quantum gravity to be able to resolve the singularity problem.

Along similar lines,  Balbinot and Poisson~\cite{93:bp} have estimated the
quantum stress-energy tensor {}from the conservation law and the trace anomaly,
on the Ori background~\cite{Ori:91}.  However, the sign of the coefficient of
the dominant term is ambiguous and does not allow a firm conclusion.  Their
results suggest that quantum effects may either act to curb or intensify mass
inflation.

Without doubt the most serious drawback in our work is the imposition (or not)
of boundary conditions.  It may, in fact, be possible to obtain some of the
nonlocal terms by examining the solution for $r^2 \gg e^2$ and matching the
solutions near $r=|e|$. Nevertheless, it is difficult to see how this will
alter our conclusions and for this reason such an effort does not seem to us
worthwhile in this case. A more complete analysis in the original mass
inflation scenario~\cite{P&I:89} is under way, and some of the deficiencies of
this model are dealt with in more detail there~\cite{IsraelAB}.

\acknowledgments

The authors would like to thank Roberto Balbinot, Andrei Barvinsky, Werner
Israel, Amos Ori, Don Page, and Eric Poisson for helpful discussion. RC
acknowledges support form the Natural Sciences and Engineering Research Council
of Canada.

\end{document}